\documentclass[aps,prl,twocolumn,groupedaddress,amsmath,amssymb]{revtex4-1}
\usepackage{url}
\usepackage{hyperref}
\usepackage{mathtools}
\usepackage{tensor}

\begin{document}

%Title of paper
\title{Poincar\'e Gauge Gravity Cosmology}

\author{Hongchao Zhang}
\email{zhanghc@mail.dlut.edu.cn}
\author{Lixin Xu}
\email{lxxu@dlut.edu.cn}

\affiliation{Institute of Theoretical Physics, School of Physics and Optoelectronic Technology, \\ Dalian University of Technology, Dalian, 116024, People's Republic of China}

\date{\today}

\begin{abstract}
% insert abstract here
In this work, we construct the logical framework of the Poincar\'e gauge gravity cosmology based on five postulations, and introduce the modified redshift relation within this framework. Then we solve a system with quadratic action and some other assumptions to get an analytic solution on background level. The evolution of the Universe on background can be reproduced from this solution without hypothesizing dark energy. Further, we use the type Ia supernova data set JLA to test the effect of the modified redshift relation under the constraints of system parameters. The results show that the constraint on some parameters are compact.
\end{abstract}

% insert suggested PACS numbers in braces on next line
\pacs{98.80.-k, 98.80.Es}

%\maketitle must follow title, authors, abstract, \pacs, and \keywords
\maketitle

% Paragraph 1
The general relativity (GR) established by Einstein one hundred years ago is a theory on gravity or spacetime.
With the continuous efforts of many scientists, both in the weak field limit (as in the Solar System) and with the stronger fields present in systems of binary pulsars the predictions of general relativity have been extremely well testified locally.
However, the modern cosmology which is based on GR, can not explain some phenomenons of the Universe on large scale, such as the galaxy rotation curves problem \cite{rubin1980rotational} and the late-time acceleration of the Universe \cite{riess1998observational} \cite{perlmutter1999measurements}.
GR is so elegant that most of the cosmologists would rather invoke dark matter (DM) and dark energy (DE) to make it self-consistently than deny it.
Nevertheless, the rest of the cosmologists try to modify it or simply find a new one.
For decades people has proposed various dark models and modified gravity theories, but no one satisfy both theoretical self-consistent and in conformity with the observations.

% Paragraph 2
Let us further revoice the issues from the perspective of logic and geometry.
In GR, the spacetime is given by a Riemannian manifold \{${\mathbf{V}}_{4}, g$\}, where g is the only structure and  the metric-compatible connection $\Gamma$ is not an independent object.
One of fundamental postulations of GR, the equivalence principle, mathematically equivalent to that the connection be symmetric in the lower indeces, namely torsion, $T$, vanished \cite{das2011lectures}.
Thus $\Gamma$ degenerates to the Levi-Civita connection constructing by $g$.
However, if we proceed from the pure geometry, as two fundamental structures on a affinely connected metric manifold, there is no reason apart from simplicity to assume any special relation between $\Gamma$ and $g$ \cite{cai2015f}.
In fact, there are three independent structures on the general affinely connected metric manifold, i.e. \{${\mathbf{L}}_{4}, g, \Gamma, Q$\}, where $Q \equiv - \nabla g$ is the non-metricity.
A manifold \{${\mathbf{L}}_{4}, g, \Gamma, Q$\} with the metric-compatiable condition is called the Riemann-Cartan manifold and denoted by \{${\mathbf{U}}_{4}, g, \Gamma$\} \cite{gronwald1996gauge}.
Furthermore, from the viewpoint of symmetry, the Poincar\'e group, as the group of Minkowski spacetime isometries, carries the full symmetry of Minkowski spacetime \cite{minkowski1910grundgleichungen}, and is the generalization of Lorentz group, the fondamental group of GR.
Localization of Poincar\'e symmetry leads to the Poincar\'e gauge theory (PGT) of gravity, which contains GR as a special case \cite{Kibble1961}.
By analyzing the local Poincar\'e gauge invariant of a Lagrangian system, one find that PGT has the geometric structure of the Riemann-Cartan manifold \{${\mathbf{U}}_{4}, g, \Gamma$\} \cite{Blagojevic2001,sciama1962analogy}.

% Paragraph 3
Here we focus on the cosmological application of PGT, and try to provide a theoretical interpretation of the late-time acceleration.
In order to make our statement more clearly, some conventions is necessary to declare before the calculations.
So we introduce five general postulations which could be regarded as the logical foundations of the Poincar\'e gauge gravity cosmology (PGGC), while some particular assumptions are proposed just for the convenience of handling.

% postulation a
\paragraph{Postulation 1. The spacetime is a Riemann-Cartan manifold \{${\mathbf{U}}_{4}, g, T$\}, where metric $g$ and torsion $T$ are viewed as two independent structures.}
While the affinely connection $\Gamma$ is not independent any more, it can be derived from $g$ and $T$ in the following way,
\begin{equation}
\Gamma^{\rho}{}_{\mu\nu} = \{^{\rho}{}_{\mu\nu}\}+\frac{1}{2} \left( T^{\rho}{}_{\mu\nu}+T_{\mu}{}^{\rho}{}_{\nu}+T_{\nu}{}^{\rho}{}_{\mu} \right),
\label{eq:GammaToT}
\end{equation}
where $\{^{\rho}{}_{\mu\nu}\}$ is the Levi-Civita connection.

% postulation b
\paragraph{Postulation 2. The contents of the Universe are spinless on cosmological scale.}
Since the spin orientation for ordinary metter is random, the macroscopic space average of the spin vanishes.
There may exist celestial bodies with a global spin, while, we do not care here.
This postulation means that the Lagrangian density of matter fields do not depend on the torsion tensor.
Thus by Rosenfeld's prescription \cite{rosenfeld1940tenseur}, the spin tensor, defined as follows,
\begin{equation}
S_{\rho}{}^{\mu\nu} \coloneqq \frac{1}{\sqrt{\vert g\vert}} \frac{\delta \left(\sqrt{\vert g\vert} \mathcal{L}_{m} \right)}{\delta T^{\rho}{}_{\mu\nu}},
\label{eq:defSpinTensor}
\end{equation}
vanishes \cite{lu2016cosmology}.

% postulation c
\paragraph{Postulation 3. Also about the contents of the Universe. No dark energy hypothesis (also cosmological constant), dark matter and baryonic matter are indistinguishable. In addition, photons and neutrinos with or without mass are under consideration.}
Therefore, the energy-momentum tensor defined at follows,
\begin{equation}
T^{\mu\nu} \coloneqq \frac{1}{\sqrt{\vert g\vert}} \frac{\delta \left(\sqrt{\vert g\vert} \mathcal{L}_{m} \right)}{\delta g_{\mu\nu}},
\label{eq:defEMTensor}
\end{equation}
is just the one which we usually used in cosmology.

% postulation d
\paragraph{Postulation 4. An action, which satisfies the Poincar\'e gauge invariance.}
This condition guarantees that the field equations and conservation laws of PGGC can be obtained from the last action principle.
However, such action has an infinite number.
Here, as an example, we adopt an action in such a formulation \cite{chee2012exact}:
\begin{eqnarray}
s &=& \int d^{4}x \sqrt{\vert g\vert}\left[\mathcal{L}_{g}+\mathcal{L}_{m} \right],\\
\mathcal{L}_{g} &\equiv& \frac{1}{2}R+\alpha R^{2}+\beta R_{\mu\nu}R^{\mu\nu}+\gamma T^{\rho}{}_{\mu\nu}T_{\rho}{}^{\mu\nu},
\label{eq:defAction}
\end{eqnarray}
where $\alpha$, $\beta$ are two parameters with the dimension of $\left[\emph{L}\right]^{2}$, and $\gamma$ is a dimensionless parameter. The terms $\frac{1}{2}R+\gamma T^{\rho}{}_{\mu\nu}T_{\rho}{}^{\mu\nu}$ represent weak gravity, while $\alpha R^{2}+\beta R_{\mu\nu}R^{\mu\nu}$ represent strong gravity. The Ricci scalar and Ricci tensor are constructed by the affinely connection $\Gamma$.

% postulation e
\paragraph{Postulation 5. The cosmological principle is adopted as usual, namely, our Universe is homogeneous and isotropic when viewed on a large enough scale.}
This postulation alone determine the spacetime metric up to the Friedmann-Robertson-Walker (FRW) form \cite{wald2010general}:
\begin{equation}
ds^{2} = -dt^{2}+a^{2}(t)
\left[ d\mathbf{x}^{2}+K \frac{\left(\mathbf{x}\cdot d\mathbf{x} \right)^{2}}{1-K \mathbf{x}^{2}} \right].
\label{eq:FRWmetric}
\end{equation}
In this work, we consider the spatially flat case, thus $K = 0$.
Meanwhile, torsion can be decomposed with respect to the Lorentz group into three irreducible tensors
\begin{equation}
T^{\rho}{}_{\mu\nu} = \prescript{V}{}T^{\rho}{}_{\mu\nu}+\prescript{A}{}T^{\rho}{}_{\mu\nu}+\prescript{T}{}T^{\rho}{}_{\mu\nu},
\label{eq:DecampositionTorsion}
\end{equation}
where $V$, $A$, $T$ in the upper left corner represent the vector (trace), the axial (totally anti-symmetric) and the tensor (traceless non totally anti-symmetric) parts respectively \cite{cai2015f}.
According to \cite{tsamparlis1981methods}, it turns out that the only torsion tensors compatible with a FRW universe are the time-like vector torsion and the time-like axial torsion, which can be constituted in terms of a scalar torsion $h(t)$ and a pseudoscalar torsion $f(t)$,
\begin{equation}
T_{ij0} = a^2 h \delta_{ij}, ~~ T_{ijk} = a^3 f \epsilon_{ijk}.
\label{eq:TorsionScalar}
\end{equation}

% Paragraph 9
\paragraph{The modified redshift relation.}
A photon propagates, in a Riemann-Cartan spacetime, along with a autoparallel curve (should be distinguished from the geodesic), satisfies the following autoparallel equation
\begin{equation}
\frac{d^2 x^{\rho}}{d\xi^2} + \Gamma^{\rho}{}_{(\mu\nu)} \frac{d x^{\mu}}{d\xi} \frac{d x^{\nu}}{d\xi} = 0,
\label{eq:Autoparallel}
\end{equation}
where $\xi$ is the affinely parameter.
It can be seen from Eq.(\ref{eq:GammaToT}) that the symmetric part of $\Gamma$ contains torsion too, thus the dispersion relations should be modified by a torsion term
\begin{equation}
\frac{\dot{E}}{E} + \frac{\dot{a}}{a} + h = 0,
\label{eq:Autoparallel}
\end{equation}
where $E$ is the energy of photon, and ``$\cdot$'' denotes the derivative in terms of $t$.
The redshift $z$, defined as the rate of wavelength change, is modified as
\begin{equation}
z \equiv \frac{\lambda_0}{\lambda}-1 = \frac{a_0}{a}\exp{\bigg[-\int^{t}_{t_0}h(\tau)d\tau \bigg]}-1.
\label{eq:Red-shift}
\end{equation}
It is reasonable to interpret this change as the influence of the internal rotation freedom of the spacetime on the photon spectroscopy.
The cosmological distances, such as the comoving distance, the luminosity distance and the angular diameter distance, which are defined based on the redshift relation, should be modified.
And some cosmological probes, which measure based on these distances, should be re-examined.

% Paragraph 10
\paragraph{The background solutions of PGGC.}
The modified Einstein equation and Cartan equation can be obtained immediately from the variational principles in terms of metric $g$ and torsion $T$ respectively.
Subsequently, we get ten component equations: two from the modified Einstein equation, five from the modified Cartan equation, one from the modified conservation Einstein equation, and two from the modified conservation Cartan equation.
To analyze these equations, it is convenient to nondimensionalize all variables and then we get a very complicated system of nonlinear algebraic equations.
It is possible to solve this nonlinear system by powerful computers.
Here we introduce the single fluid assumption to simplify the calculation, i.e. in each period of time (before and after the epoch of matter-radiation equality) we regard the universe to be dominated by a single fluid.
Then make use of some mathematical techniques, we find one set of solutions.
After completing the dimensions, these solutions become differential equations again.
The analytic solutions can be easily obtained after setting the only initial condition of Hubble, $H(a_{0}=1) = H_{0}$.
The final solutions of PGGC on background level are

% hm(a)
\begin{equation}
h_{m}(a) = \frac{\sqrt{Cmc}}{\sqrt{Cmb}} \sqrt{(Cme+1) a^{\frac{1}{Cmd}}-1},
\label{eq:hm}
\end{equation}
% rhom(a)
\begin{equation}
8\pi G \rho_{m}(a) = Cmf + Cmg \left[(Cme+1) a^{\frac{1}{Cmd}}-1\right],
\label{eq:rhom}
\end{equation}
% Hm(a)
\begin{equation}
H_{m}(a) = -\frac{2 \sqrt{Cmc} \sqrt{Cmb} Cmd} {Cma} \sqrt{(Cme+1)
   a^{\frac{1}{Cmd}}-1},
\label{eq:Hm}
\end{equation}

\begin{widetext}
% hr(a)
\begin{equation}
h_{r}(a) = \frac{\sqrt{Crc}}{\sqrt{Crb}}
   \sqrt{
   \left\{\frac{Cmc Crb}{Cmb Crc} \left[(Cme+1)
  a_{eq}^{\frac{1}{Cmd}}-1\right]+1\right\} \left(\frac{a}{a_{eq}}\right)^{\frac{1}{Crd}}-1},
\label{eq:hr}
\end{equation}
% rhor(a)
\begin{equation}
8\pi G \rho_{r}(a) = Crf + Crg \bigg\{
   \left\{\frac{Cmc Crb }{Cmb Crc}\left[(Cme+1)
   a_{eq}^{\frac{1}{Cmd}}-1\right]+1\right\} \left(\frac{a}{a_{eq}}\right)^{\frac{1}{Crd}}-1\bigg\},
\label{eq:rhor}
\end{equation}
% Hr(a)
\begin{equation}
H_{r}(a) = -\frac{2 \sqrt{Crc} \sqrt{Crb} Crd}{Cra}
   \sqrt{
   \left\{\frac{Cmc Crb}{Cmb Crc} \left[(Cme+1)
  a_{eq}^{\frac{1}{Cmd}}-1\right]+1\right\} \left(\frac{a}{a_{eq}}\right)^{\frac{1}{Crd}}-1},
\label{eq:Hr}
\end{equation}
\end{widetext}
where $Cma \sim Cmg$, $Cra \sim Crg$ are coefficients consisting of $\alpha, \beta, \gamma$ and $H_{0}$.
$m$ and $r$ denote objects in the matter-dominated and the radiation-dominated era respectively.
$Cmf = Crf$ can be regarded as the ``cosmological constant'' times $8 \pi G$ formally.
$H(a) \equiv \dot{a}/a$ is the Hubble rate.
Eq.(\ref{eq:hm})(\ref{eq:Hm}) and Eq.(\ref{eq:hr})(\ref{eq:Hr}) show that $h(a) \propto H(a)$, which is consistent with Eq.(\ref{eq:Autoparallel}).
The function of redshift $z(a)$ can be written from Eq.(\ref{eq:Red-shift}) immediately
\begin{equation}
z_{m}(a) = a^{\frac{Cma}{2 Cmb Cmd}-1} -1,
\label{zofam}
\end{equation}
\begin{equation}
z_{r}(a) = a_{eq}^{\frac{Cma}{2 Cmb Cmd}-\frac{Cra}{2 Crb Crd}} a^{\frac{Cra}{2 Crb Crd}-1} -1,
\label{zofar}
\end{equation}
where $\frac{Cma}{2 Cmb Cmd}$ and $\frac{Cra}{2 Crb Crd}$ are the modified terms come from torsion.

\begin{figure*}[tbp]
\includegraphics[width=1.\textwidth]{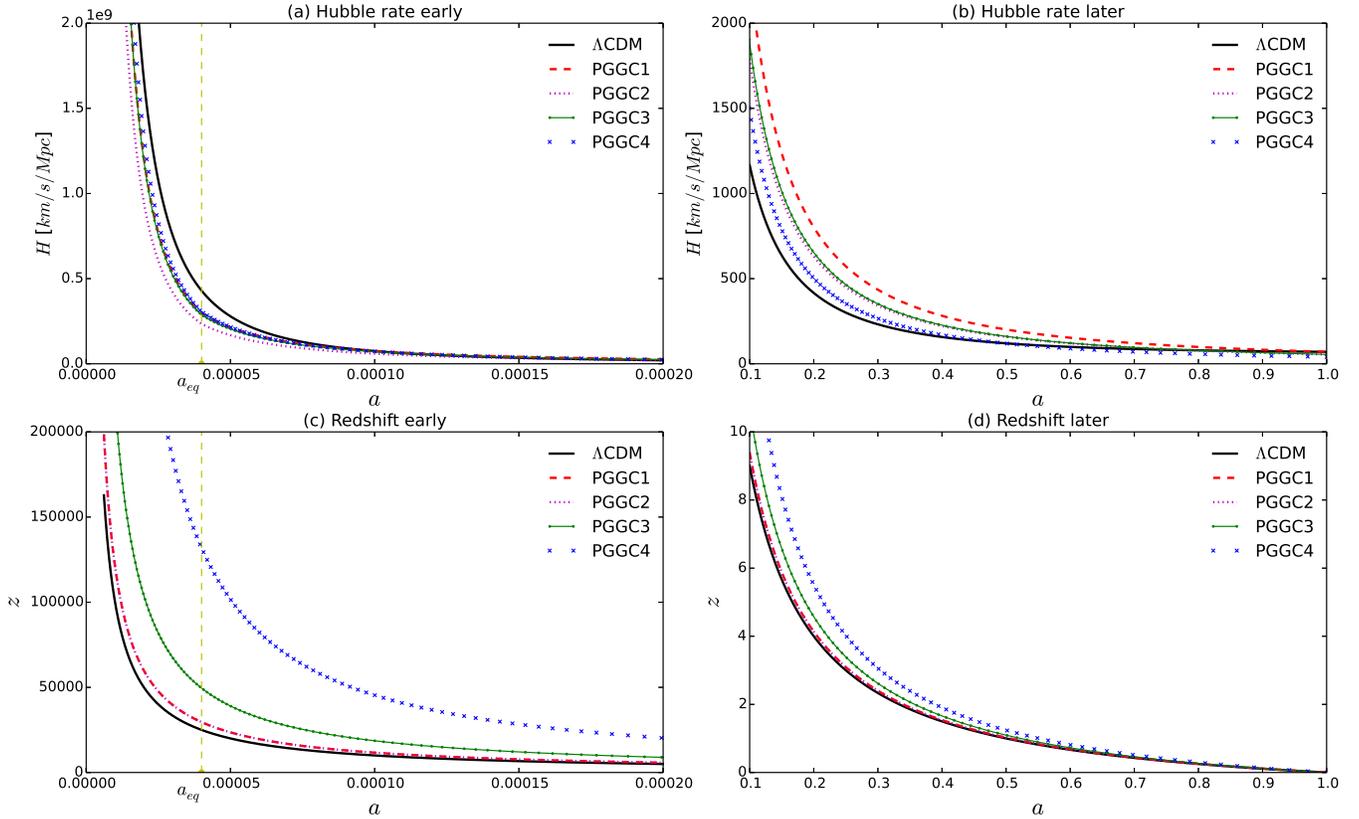}%
\caption{\label{fig:Hubble_redshift} The comparison among five parameterized PGGC models and the $\Lambda$CDM, where the values of parameters can be found in TABLE \ref{tab:PGGCvalues}. Panels (a), (b): The Hubble rate in early and later period. Panels (c), (d): The redshift in early and later period, where the formula of redshift in $\Lambda$CDM is $\frac{1}{a}-1$. In panels (a), (c), the epoch of matter-radiation equality i.e. $a_{eq}$ has been denoted by a vertically dashed line.}
\end{figure*}

\begin{figure}[htbp]
\includegraphics[width=.5\textwidth]{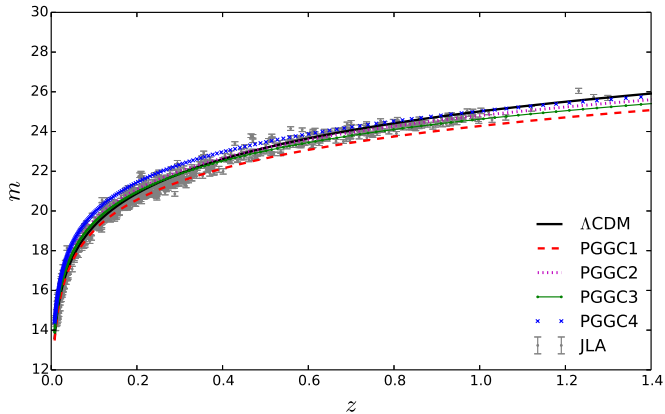}%
\caption{\label{fig:magnitude} The theoretical apparent magnitude $m$ versus the redshift $z$ for $\Lambda$CDM and PGGCs. The scattered points with error bars represent the distribution of apparent magnitude of $740$ SNIa from the JLA.}
\end{figure}

% Paragraph 11
\paragraph{Comparing with the $\Lambda$CDM model.}
In order to assure that $Cmb > 0$, $Cmc > 0$, and $Cma Cmd < 0$, we get a range of the estimates of $\alpha$, $\beta$, $\gamma$,
\begin{equation}
\alpha \leqslant 0 ~and~ \beta < 20 \alpha ~and~ \gamma > \frac{5\beta}{-80\alpha+4\beta}.
\label{eq:Range}
\end{equation}
Indeed, the solutions of this nonlinear system of inequalities are very complicated.
However, this simple solution show us a direction to go forward.
We have said that $\alpha$, $\beta$ possess the dimension of $\left[\emph{L}\right]^{2}$.
It is convenient to choose a quantity with such a dimension in cosmology as the unit of $\alpha$ and $\beta$, that is, $3H^{2}_{0}/c^{2}$, which is denoted as $grhom$ in \texttt{camb} \cite{lewis2000efficient}.
To make a comparison among the PGGC models and the $\Lambda$CDM, we plot the curves of the Hubble rate and the redshift in early and later period respectively in FIG. \ref{fig:Hubble_redshift}.
We fix the value of $\alpha$ on $-1 grhom$ and $a_{eq}$ on $0.00004$.
Curves PGGC1 to PGGC4 show the influence of the remanent parameters, where the values of $\beta$, $\gamma$, $H_{0}$ are listed in TABLE \ref{tab:PGGCvalues}.
In addition the values of parameters in $\Lambda$CDM are set to: $\Omega_{m}=0.0419$, $\Omega_{c}=0.239$, $\Omega_{r}=0.00123$, $\Omega_{\Lambda}=0.718$, and $H_{0}=72.1 ~km/s/Mpc$.
From those curves we can see that it is possible to reproduce the tendency of $\Lambda$CDM from PGGC without the assumption of dark energy through choosing appropriate value of parameters.
If there exists a type Ia supernova (SNIa) with a certain redshift, it can be inferred from panel (d) that the corresponding scale factor $a$ of this supernova maybe very different in $\Lambda$CDM and PGGCs, however according to panel (b), it is possible to get the conclusion that the Hubble rates in this position obtained from different theories are exactly equivalent.
To show this in an absolute way, we plot the theoretical apparent magnitude $m$ versus the redshift $z$ for $\Lambda$CDM and PGGCs in FIG. \ref{fig:magnitude}, and as a comparison, we plot the distribution of apparent magnitude of $740$ SNIa from the Joint Light-curve Analysis (JLA) sample \cite{betoule2014improved}.
Where the relation between the apparent magnitude $m$ and the luminosity distance $d_{L}$ is
\begin{equation}
m-M=5 ~log_{10}d_{L}+25.
\label{magnitude}
\end{equation}
$M$ is the absolute magnitude and we set $M=19.15$ here.
From FIG. \ref{fig:magnitude} we can see that the SNIa data points from JLA can be fitted very well by $\Lambda$CDM, PGGC2 and PGGC3.

\begingroup
\begin{center}
\begin{table}
\begin{ruledtabular}
\renewcommand{\arraystretch}{1.5}
\begin{tabular}{cccc}
%\hline\hline
Parameters & $\beta ~[grhom]$ & $\gamma$ & $H_{0} ~[km/s/Mpc]$ \\ \hline
PGGC1 & $-30$ & $4$ & $70$\\
PGGC2 & $-30$ & $4$ & $55$\\
PGGC3 & $-30$ & $5$ & $55$\\
PGGC4 & $-40$ & $4$ & $40$\\
%\hline\hline
\end{tabular}
\caption{The values of $\beta$, $\gamma$, $H_{0}$ for curves PGGC1 to PGGC4 in FIG. \ref{fig:Hubble_redshift}, where we have fixed $\alpha$ on $-1 grhom$ and $a_{eq}$ on $0.00004$.}\label{tab:PGGCvalues}
\end{ruledtabular}
\end{table}
\end{center}
\endgroup

% Paragraph 12
\paragraph{Constraint from SNIa data.}
The type Ia supernova observation is a smoking gun for the late-time cosmic acceleration.
It is a probe on background level and simple enough to check the effect from the modifies redshift relation and Hubble rate through luminosity distance which we have seen in FIG. \ref{fig:magnitude}.
Here we use the SNIa from JLA to constrain the parameters, $\alpha$, $\beta$, $\gamma$.
Since the redshift of supernovas from JLA distribute less than $1.5$, we are not confident in getting good results and it is obvious that the ability to constrain $a_{eq}$ is poor.
It is impossible to constrain $H_{0}$ by SNIa because the $H_{0}$ can be eliminated in the $\chi^{2}$.
We modify the \texttt{CosmoMC} code \cite{lewis2002cosmological} to include the PGGC model.

\begingroup                                                                                                                     
\squeezetable                                                                                                                   
\begin{center}                                                                                                                  
\begin{table}
\begin{ruledtabular}
\renewcommand{\arraystretch}{1.6}                                           
\begin{tabular}{cccc}                                                                                                            
%\hline\hline                                                                                                                    
Parameters & Priors & Mean with errors & Best fit \\ \hline
$\alpha$ & $\left[-5, 5\right]$ & $   -1.70_{-    1.08-    1.93-    2.04}^{+    0.839+    1.92+    3.41}$ & $   -0.939$\\
$\beta$ & $\left[-200, 10\right]$ & $ -139_{-   60.0-   60.0-   60.0}^{+   17.9+   82.2+  106}$ & $ -110$\\
$\gamma$ & $\left[0, 50\right]$ & $   24.1_{-   20.3-   24.1-   24.1}^{+   11.3+   25.8+   25.8}$ & $    6.02$\\
$H_{0}$ & $70$ &  - &  -\\
$a_{eq}$ & $0.00004$ &  - &  -\\
\hline
$\alpha_{JLA}$ &-& $  0.141_{-  0.00673-  0.0133-  0.0169}^{+  0.00674+  0.0134+  0.0175}$ & $  0.140$\\
$\beta_{JLA}$ &-& $    3.10_{-    0.0797-    0.156-    0.198}^{+    0.0802+    0.161+    0.214}$ & $    3.10$\\
$\chi^2_{\rm JLA}$ &-& $  698_{-    3.01-    4.14-    4.62}^{+    1.18+    5.37+    9.15}$ & $  695$\\
%\hline\hline                                                                                                                    
\end{tabular}                                                                                                                   
\caption{The priors, mean values with $1\sigma$, $2\sigma$, $3\sigma$ limits and the best fit values for the PGGC and SNIa parameters.}\label{tab:table_6107}
\end{ruledtabular}                                                                                                   
\end{table}                                                                                                                     
\end{center}                                                                                                                    
\endgroup

\begin{figure}[htbp]
\includegraphics[width=.5\textwidth]{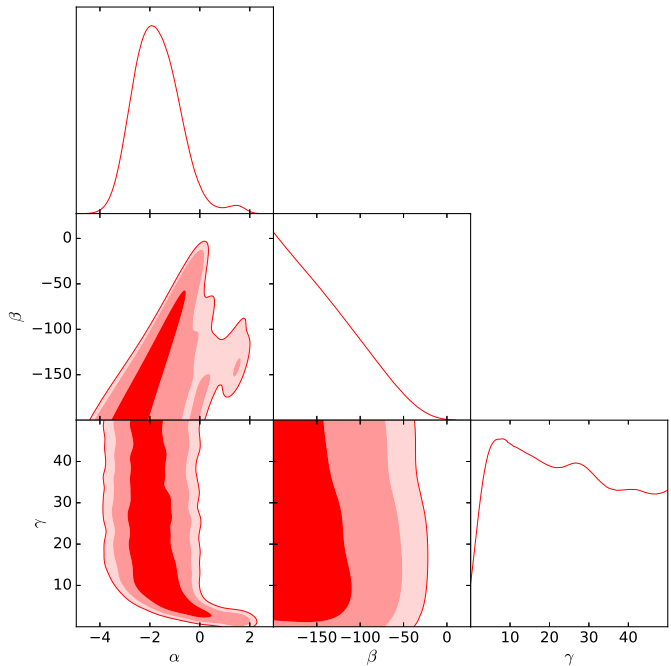}%
\caption{\label{fig:contour_6112} The $1D$ marginalized distribution and $2D$ contours for $\alpha$, $\beta$, $\gamma$ with $1\sigma$, $2\sigma$, $3\sigma$ confidence levels.}
\end{figure}

\begin{acknowledgments}
% put your acknowledgments here.
Lixin Xu is supported in part by National Natural Science Foundation of China under Grant No. 11275035, Grant No.
11675032 (People's Republic of China), and supported by ``the Fundamental Research Funds for the Central Universities'' under Grant No. DUT16LK31.
\end{acknowledgments}

% Create the reference section using BibTeX:
\bibliography{PGGC_article}

%merlin.mbs apsrev4-1.bst 2010-07-25 4.21a (PWD, AO, DPC) hacked
%Control: key (0)
%Control: author (8) initials jnrlst
%Control: editor formatted (1) identically to author
%Control: production of article title (-1) disabled
%Control: page (0) single
%Control: year (1) truncated
%Control: production of eprint (0) enabled
\begin{thebibliography}{18}%
\makeatletter
\providecommand \@ifxundefined [1]{%
 \@ifx{#1\undefined}
}%
\providecommand \@ifnum [1]{%
 \ifnum #1\expandafter \@firstoftwo
 \else \expandafter \@secondoftwo
 \fi
}%
\providecommand \@ifx [1]{%
 \ifx #1\expandafter \@firstoftwo
 \else \expandafter \@secondoftwo
 \fi
}%
\providecommand \natexlab [1]{#1}%
\providecommand \enquote  [1]{``#1''}%
\providecommand \bibnamefont  [1]{#1}%
\providecommand \bibfnamefont [1]{#1}%
\providecommand \citenamefont [1]{#1}%
\providecommand \href@noop [0]{\@secondoftwo}%
\providecommand \href [0]{\begingroup \@sanitize@url \@href}%
\providecommand \@href[1]{\@@startlink{#1}\@@href}%
\providecommand \@@href[1]{\endgroup#1\@@endlink}%
\providecommand \@sanitize@url [0]{\catcode `\\12\catcode `\$12\catcode
  `\&12\catcode `\#12\catcode `\^12\catcode `\_12\catcode `\%12\relax}%
\providecommand \@@startlink[1]{}%
\providecommand \@@endlink[0]{}%
\providecommand \url  [0]{\begingroup\@sanitize@url \@url }%
\providecommand \@url [1]{\endgroup\@href {#1}{\urlprefix }}%
\providecommand \urlprefix  [0]{URL }%
\providecommand \Eprint [0]{\href }%
\providecommand \doibase [0]{http://dx.doi.org/}%
\providecommand \selectlanguage [0]{\@gobble}%
\providecommand \bibinfo  [0]{\@secondoftwo}%
\providecommand \bibfield  [0]{\@secondoftwo}%
\providecommand \translation [1]{[#1]}%
\providecommand \BibitemOpen [0]{}%
\providecommand \bibitemStop [0]{}%
\providecommand \bibitemNoStop [0]{.\EOS\space}%
\providecommand \EOS [0]{\spacefactor3000\relax}%
\providecommand \BibitemShut  [1]{\csname bibitem#1\endcsname}%
\let\auto@bib@innerbib\@empty
%</preamble>
\bibitem [{\citenamefont {Rubin}\ \emph {et~al.}(1980)\citenamefont {Rubin},
  \citenamefont {Ford~Jr},\ and\ \citenamefont
  {Thonnard}}]{rubin1980rotational}%
  \BibitemOpen
  \bibfield  {author} {\bibinfo {author} {\bibfnamefont {V.~C.}\ \bibnamefont
  {Rubin}}, \bibinfo {author} {\bibfnamefont {W.~K.}\ \bibnamefont {Ford~Jr}},
  \ and\ \bibinfo {author} {\bibfnamefont {N.}~\bibnamefont {Thonnard}},\
  }\href@noop {} {\bibfield  {journal} {\bibinfo  {journal} {The Astrophysical
  Journal}\ }\textbf {\bibinfo {volume} {238}},\ \bibinfo {pages} {471}
  (\bibinfo {year} {1980})}\BibitemShut {NoStop}%
\bibitem [{\citenamefont {Riess}\ \emph {et~al.}(1998)\citenamefont {Riess},
  \citenamefont {Filippenko}, \citenamefont {Challis}, \citenamefont
  {Clocchiatti}, \citenamefont {Diercks}, \citenamefont {Garnavich},
  \citenamefont {Gilliland}, \citenamefont {Hogan}, \citenamefont {Jha},
  \citenamefont {Kirshner} \emph {et~al.}}]{riess1998observational}%
  \BibitemOpen
  \bibfield  {author} {\bibinfo {author} {\bibfnamefont {A.~G.}\ \bibnamefont
  {Riess}}, \bibinfo {author} {\bibfnamefont {A.~V.}\ \bibnamefont
  {Filippenko}}, \bibinfo {author} {\bibfnamefont {P.}~\bibnamefont {Challis}},
  \bibinfo {author} {\bibfnamefont {A.}~\bibnamefont {Clocchiatti}}, \bibinfo
  {author} {\bibfnamefont {A.}~\bibnamefont {Diercks}}, \bibinfo {author}
  {\bibfnamefont {P.~M.}\ \bibnamefont {Garnavich}}, \bibinfo {author}
  {\bibfnamefont {R.~L.}\ \bibnamefont {Gilliland}}, \bibinfo {author}
  {\bibfnamefont {C.~J.}\ \bibnamefont {Hogan}}, \bibinfo {author}
  {\bibfnamefont {S.}~\bibnamefont {Jha}}, \bibinfo {author} {\bibfnamefont
  {R.~P.}\ \bibnamefont {Kirshner}},  \emph {et~al.},\ }\href@noop {}
  {\bibfield  {journal} {\bibinfo  {journal} {The Astronomical Journal}\
  }\textbf {\bibinfo {volume} {116}},\ \bibinfo {pages} {1009} (\bibinfo {year}
  {1998})}\BibitemShut {NoStop}%
\bibitem [{\citenamefont {Perlmutter}\ \emph {et~al.}(1999)\citenamefont
  {Perlmutter}, \citenamefont {Aldering}, \citenamefont {Goldhaber},
  \citenamefont {Knop}, \citenamefont {Nugent}, \citenamefont {Castro},
  \citenamefont {Deustua}, \citenamefont {Fabbro}, \citenamefont {Goobar},
  \citenamefont {Groom} \emph {et~al.}}]{perlmutter1999measurements}%
  \BibitemOpen
  \bibfield  {author} {\bibinfo {author} {\bibfnamefont {S.}~\bibnamefont
  {Perlmutter}}, \bibinfo {author} {\bibfnamefont {G.}~\bibnamefont
  {Aldering}}, \bibinfo {author} {\bibfnamefont {G.}~\bibnamefont {Goldhaber}},
  \bibinfo {author} {\bibfnamefont {R.}~\bibnamefont {Knop}}, \bibinfo {author}
  {\bibfnamefont {P.}~\bibnamefont {Nugent}}, \bibinfo {author} {\bibfnamefont
  {P.}~\bibnamefont {Castro}}, \bibinfo {author} {\bibfnamefont
  {S.}~\bibnamefont {Deustua}}, \bibinfo {author} {\bibfnamefont
  {S.}~\bibnamefont {Fabbro}}, \bibinfo {author} {\bibfnamefont
  {A.}~\bibnamefont {Goobar}}, \bibinfo {author} {\bibfnamefont
  {D.}~\bibnamefont {Groom}},  \emph {et~al.},\ }\href@noop {} {\bibfield
  {journal} {\bibinfo  {journal} {The Astrophysical Journal}\ }\textbf
  {\bibinfo {volume} {517}},\ \bibinfo {pages} {565} (\bibinfo {year}
  {1999})}\BibitemShut {NoStop}%
\bibitem [{\citenamefont {Das}(2011)}]{das2011lectures}%
  \BibitemOpen
  \bibfield  {author} {\bibinfo {author} {\bibfnamefont {A.}~\bibnamefont
  {Das}},\ }\href@noop {} {\emph {\bibinfo {title} {Lectures on gravitation}}}\
  (\bibinfo  {publisher} {World scientific},\ \bibinfo {year}
  {2011})\BibitemShut {NoStop}%
\bibitem [{\citenamefont {Cai}\ \emph {et~al.}(2015)\citenamefont {Cai},
  \citenamefont {Capozziello}, \citenamefont {De~Laurentis},\ and\
  \citenamefont {Saridakis}}]{cai2015f}%
  \BibitemOpen
  \bibfield  {author} {\bibinfo {author} {\bibfnamefont {Y.-F.}\ \bibnamefont
  {Cai}}, \bibinfo {author} {\bibfnamefont {S.}~\bibnamefont {Capozziello}},
  \bibinfo {author} {\bibfnamefont {M.}~\bibnamefont {De~Laurentis}}, \ and\
  \bibinfo {author} {\bibfnamefont {E.~N.}\ \bibnamefont {Saridakis}},\
  }\href@noop {} {\bibfield  {journal} {\bibinfo  {journal} {arXiv preprint
  arXiv:1511.07586}\ } (\bibinfo {year} {2015})}\BibitemShut {NoStop}%
\bibitem [{\citenamefont {Gronwald}\ and\ \citenamefont
  {Hehl}(1996)}]{gronwald1996gauge}%
  \BibitemOpen
  \bibfield  {author} {\bibinfo {author} {\bibfnamefont {F.}~\bibnamefont
  {Gronwald}}\ and\ \bibinfo {author} {\bibfnamefont {F.~W.}\ \bibnamefont
  {Hehl}},\ }\href@noop {} {\bibfield  {journal} {\bibinfo  {journal} {arXiv
  preprint gr-qc/9602013}\ } (\bibinfo {year} {1996})}\BibitemShut {NoStop}%
\bibitem [{\citenamefont {Minkowski}(1910)}]{minkowski1910grundgleichungen}%
  \BibitemOpen
  \bibfield  {author} {\bibinfo {author} {\bibfnamefont {H.}~\bibnamefont
  {Minkowski}},\ }\href@noop {} {\bibfield  {journal} {\bibinfo  {journal}
  {Mathematische Annalen}\ }\textbf {\bibinfo {volume} {68}},\ \bibinfo {pages}
  {472} (\bibinfo {year} {1910})}\BibitemShut {NoStop}%
\bibitem [{\citenamefont {Kibble}(1961)}]{Kibble1961}%
  \BibitemOpen
  \bibfield  {author} {\bibinfo {author} {\bibfnamefont {T.~W.~B.}\
  \bibnamefont {Kibble}},\ }\href@noop {} {\bibfield  {journal} {\bibinfo
  {journal} {Journal of Mathematical Physics}\ }\textbf {\bibinfo {volume}
  {2(2)}},\ \bibinfo {pages} {212} (\bibinfo {year} {1961})}\BibitemShut
  {NoStop}%
\bibitem [{\citenamefont {Blagojevic}(2001)}]{Blagojevic2001}%
  \BibitemOpen
  \bibfield  {author} {\bibinfo {author} {\bibfnamefont {M.}~\bibnamefont
  {Blagojevic}},\ }\href@noop {} {\emph {\bibinfo {title} {Gravitation and
  gauge symmetries}}}\ (\bibinfo  {publisher} {CRC Press},\ \bibinfo {year}
  {2001})\BibitemShut {NoStop}%
\bibitem [{\citenamefont {Sciama}(1962)}]{sciama1962analogy}%
  \BibitemOpen
  \bibfield  {author} {\bibinfo {author} {\bibfnamefont {D.~W.}\ \bibnamefont
  {Sciama}},\ }\href@noop {} {\bibfield  {journal} {\bibinfo  {journal} {Recent
  developments in general relativity}\ ,\ \bibinfo {pages} {415}} (\bibinfo
  {year} {1962})}\BibitemShut {NoStop}%
\bibitem [{\citenamefont {Rosenfeld}(1940)}]{rosenfeld1940tenseur}%
  \BibitemOpen
  \bibfield  {author} {\bibinfo {author} {\bibfnamefont {L.~J. H.~C.}\
  \bibnamefont {Rosenfeld}},\ }\href@noop {} {\emph {\bibinfo {title} {Sur le
  tenseur d'impulsion-{\'e}nergie}}}\ (\bibinfo  {publisher} {Palais des
  Acad{\'e}mies},\ \bibinfo {year} {1940})\BibitemShut {NoStop}%
\bibitem [{\citenamefont {Lu}\ and\ \citenamefont
  {Chee}(2016)}]{lu2016cosmology}%
  \BibitemOpen
  \bibfield  {author} {\bibinfo {author} {\bibfnamefont {J.}~\bibnamefont
  {Lu}}\ and\ \bibinfo {author} {\bibfnamefont {G.}~\bibnamefont {Chee}},\
  }\href@noop {} {\bibfield  {journal} {\bibinfo  {journal} {Journal of High
  Energy Physics}\ }\textbf {\bibinfo {volume} {5}} (\bibinfo {year}
  {2016})}\BibitemShut {NoStop}%
\bibitem [{\citenamefont {Chee}\ and\ \citenamefont
  {Guo}(2012)}]{chee2012exact}%
  \BibitemOpen
  \bibfield  {author} {\bibinfo {author} {\bibfnamefont {G.}~\bibnamefont
  {Chee}}\ and\ \bibinfo {author} {\bibfnamefont {Y.}~\bibnamefont {Guo}},\
  }\href@noop {} {\bibfield  {journal} {\bibinfo  {journal} {Classical and
  Quantum Gravity}\ }\textbf {\bibinfo {volume} {29}},\ \bibinfo {pages}
  {235022} (\bibinfo {year} {2012})}\BibitemShut {NoStop}%
\bibitem [{\citenamefont {Wald}(2010)}]{wald2010general}%
  \BibitemOpen
  \bibfield  {author} {\bibinfo {author} {\bibfnamefont {R.~M.}\ \bibnamefont
  {Wald}},\ }\href@noop {} {\emph {\bibinfo {title} {General relativity}}}\
  (\bibinfo  {publisher} {University of Chicago press},\ \bibinfo {year}
  {2010})\BibitemShut {NoStop}%
\bibitem [{\citenamefont {Tsamparlis}(1981)}]{tsamparlis1981methods}%
  \BibitemOpen
  \bibfield  {author} {\bibinfo {author} {\bibfnamefont {M.}~\bibnamefont
  {Tsamparlis}},\ }\href@noop {} {\bibfield  {journal} {\bibinfo  {journal}
  {Physical Review D}\ }\textbf {\bibinfo {volume} {24}},\ \bibinfo {pages}
  {1451} (\bibinfo {year} {1981})}\BibitemShut {NoStop}%
\bibitem [{\citenamefont {Lewis}\ \emph {et~al.}(2000)\citenamefont {Lewis},
  \citenamefont {Challinor},\ and\ \citenamefont
  {Lasenby}}]{lewis2000efficient}%
  \BibitemOpen
  \bibfield  {author} {\bibinfo {author} {\bibfnamefont {A.}~\bibnamefont
  {Lewis}}, \bibinfo {author} {\bibfnamefont {A.}~\bibnamefont {Challinor}}, \
  and\ \bibinfo {author} {\bibfnamefont {A.}~\bibnamefont {Lasenby}},\
  }\href@noop {} {\bibfield  {journal} {\bibinfo  {journal} {The Astrophysical
  Journal}\ }\textbf {\bibinfo {volume} {538}},\ \bibinfo {pages} {473}
  (\bibinfo {year} {2000})}\BibitemShut {NoStop}%
\bibitem [{\citenamefont {Betoule}\ \emph {et~al.}(2014)\citenamefont
  {Betoule}, \citenamefont {Kessler}, \citenamefont {Guy}, \citenamefont
  {Mosher}, \citenamefont {Hardin}, \citenamefont {Biswas}, \citenamefont
  {Astier}, \citenamefont {El-Hage}, \citenamefont {Konig}, \citenamefont
  {Kuhlmann} \emph {et~al.}}]{betoule2014improved}%
  \BibitemOpen
  \bibfield  {author} {\bibinfo {author} {\bibfnamefont {M.}~\bibnamefont
  {Betoule}}, \bibinfo {author} {\bibfnamefont {R.}~\bibnamefont {Kessler}},
  \bibinfo {author} {\bibfnamefont {J.}~\bibnamefont {Guy}}, \bibinfo {author}
  {\bibfnamefont {J.}~\bibnamefont {Mosher}}, \bibinfo {author} {\bibfnamefont
  {D.}~\bibnamefont {Hardin}}, \bibinfo {author} {\bibfnamefont
  {R.}~\bibnamefont {Biswas}}, \bibinfo {author} {\bibfnamefont
  {P.}~\bibnamefont {Astier}}, \bibinfo {author} {\bibfnamefont
  {P.}~\bibnamefont {El-Hage}}, \bibinfo {author} {\bibfnamefont
  {M.}~\bibnamefont {Konig}}, \bibinfo {author} {\bibfnamefont
  {S.}~\bibnamefont {Kuhlmann}},  \emph {et~al.},\ }\href@noop {} {\bibfield
  {journal} {\bibinfo  {journal} {Astronomy \&amp; Astrophysics}\ }\textbf
  {\bibinfo {volume} {568}},\ \bibinfo {pages} {A22} (\bibinfo {year}
  {2014})}\BibitemShut {NoStop}%
\bibitem [{\citenamefont {Lewis}\ and\ \citenamefont
  {Bridle}(2002)}]{lewis2002cosmological}%
  \BibitemOpen
  \bibfield  {author} {\bibinfo {author} {\bibfnamefont {A.}~\bibnamefont
  {Lewis}}\ and\ \bibinfo {author} {\bibfnamefont {S.}~\bibnamefont {Bridle}},\
  }\href@noop {} {\bibfield  {journal} {\bibinfo  {journal} {Physical Review
  D}\ }\textbf {\bibinfo {volume} {66}},\ \bibinfo {pages} {103511} (\bibinfo
  {year} {2002})}\BibitemShut {NoStop}%
\end{thebibliography}%

\end{document}